\documentclass[preprint,nofootinbib,aps,superscriptaddress,floatfix]{revtex4}

\usepackage{graphicx}

\def\OMIT#1{{}}

\newcommand{\Dbar}{\,\overline{\!D}}
\newcommand{\Ksp}{K^{*+}}
\newcommand{\Ksm}{K^{*-}}
\def\D0bar{\Dbar{}^0}

\newcommand{\nn}{\nonumber}
\newcommand{\beq}{\begin{equation}}
\newcommand{\eeq}{\end{equation}}
\newcommand{\beqa}{\begin{eqnarray}}
\newcommand{\eeqa}{\end{eqnarray}}

\def\sex{S_{\rm ex}}
\def\ssign{S_{\rm sign}}
\def\spi{S_\pi}

\begin{document}

\preprint{\vbox{\hbox{LBNL-51630} \hbox{CSUHEP 02-02} \hbox{hep-ph/0210433}}}

\title{\boldmath Measuring $\gamma$ in $B^\pm \to K^\pm (K K^*)_D$ decays
\vspace*{8pt}}

\vspace*{1cm}

\author{Yuval Grossman}\email{yuvalg@physics.technion.ac.il}
\affiliation{Department of Physics, Technion--Israel Institute of Technology,\\
	Technion City, 32000 Haifa, Israel\vspace{6pt} }

\author{Zoltan Ligeti}\email{zligeti@lbl.gov}
\affiliation{Ernest Orlando Lawrence Berkeley National Laboratory,
  University of California, Berkeley, CA 94720\vspace{6pt} }

\author{Abner Soffer}\email{abi@slac.stanford.edu}
\affiliation{Department of Physics, Colorado State University, 
  Fort Collins, CO 80523 \\[20pt] $\phantom{}$ }

\begin{abstract} \vspace*{8pt}

We develop a method to measure the CKM angle $\gamma$ without hadronic
uncertainties from the analysis of $B^\pm \to K^\pm D^0$ and $K^\pm
\D0bar$ followed by singly Cabibbo-suppressed $D$ decays to non
CP-eigenstates, such as $K^\pm K^{*\mp}$.  This method utilizes the
interference between $b\to c\bar u s$ and $b\to u\bar c s$ decays, and
we point out several attractive features of it.  All the modes that
need to be measured for this method are accessible in the present data.

\end{abstract}

\maketitle


Some of the theoretically cleanest determinations of the weak phase $\gamma$
rely on $B\to D K$ and related decays~\cite{GL,GW,review,book} (for
definitions, see~\cite{book,review}).  The original idea of Gronau and Wyler
(GW)~\cite{GW} was to measure two decay rates arising from $b \to c \bar u s$
and $b \to u \bar c s$ amplitudes.  By measuring the rate of a third decay that
involves the interference between these two amplitudes, one can gain
sensitivity to their relative phase, which is $\gamma$.  Since all the quarks
which appear in $B\to D K$ decays have distinct flavors, the theoretical
uncertainty arises only from higher order weak interaction effects (including,
possibly, $D-\Dbar$ mixing, which we discuss below).  However, there are no
penguin contributions to these decays.

A practical difficulty of the GW method is that the amplitude ratio $A(B^-\to
\D0bar K^-) /\break A(B^-\to D^0 K^-)$ is expected to be small. As a result,
the measurement of $|A(B^-\to \D0bar K^-)|$ using hadronic $D$ decays is
hampered by a significant contribution from the decay $B^-\to D^0 K^-$,
followed by a doubly Cabibbo-suppressed decay of the $D^0$.  To avoid this
problem, Atwood, Dunietz, and Soni (ADS)~\cite{ADS} proposed to study final
states where Cabibbo-allowed and doubly Cabibbo-suppressed $D$ decays
interfere. Several other variants of the GW method have been 
proposed~\cite{other,APS}. 
An important point is that most of the methods require the
measurements of very small rates, which have yet to be observed.
(One exception is Ref.~\cite{APS}, where relatively large rates are expected.)

In this letter we propose to use singly Cabibbo-suppressed $D$ decays to final
states that are not CP eigenstates, and have sizable rates in both $D^0$ and
$\D0bar$ decay. The use of such final states has been mentioned in 
Ref.~\cite{ADS}; here we develop the details and point out the advantages.
The simplest example is the final states $\Ksp K^-$ and
$\Ksm K^+$.  In the formalism below, we assume that all relevant decays are
dominated by the standard model amplitudes.\footnote{New physics would have to
compete with tree-level $b\to u\bar c s$ or $c\to s\bar s u$ decays, or give
rise to $D-\Dbar$ mixing near the present experimental limits to influence our
determination of $\gamma$.}  We define
\beq
A_B \equiv A(B^- \to D^0 K^-)\,, \qquad
  \bar A_B \equiv A(B^- \to \D0bar K^-)\,.
\eeq
We shall further denote
\beq
{\bar A_B\over A_B} = r_B\, e^{i (\delta_B - \gamma)}\,, \qquad
  r_B \equiv \left|{\bar A_B \over A_B}\right|,
\eeq
where $\delta_B$ is the relative strong phase between $\bar A_B$ and $A_B$, and
we have neglected the deviation of the weak phase of $\bar A_B / A_B$ from
$-\gamma$, which is suppressed by four powers of the Cabibbo angle ($\lambda^4
\simeq 2 \times 10^{-3}$).  Then the ratio of the CP conjugate decay amplitudes
is given by
\beq\label{Bratio}
{A(B^+ \to D^0 K^+) \over A(B^+ \to \D0bar K^+)}
  = r_B\, e^{i (\delta_B+\gamma)}\,. \qquad
\eeq
We denote the following $D$ decay amplitudes
\beq
A_D \equiv A(D^0 \to \Ksp K^-)\,, \qquad 
  \bar A_D \equiv A(\D0bar \to \Ksp K^-)\,,
\eeq
and their ratio
\beq
{\bar A_D\over A_D} = r_D\, e^{i \delta_D}\,, \qquad
  r_D \equiv \left|{\bar A_D \over A_D}\right|.
\eeq
Here we neglected the $c\to u$ penguin contribution compared to the $c\to s
\bar s u$ tree diagram, which is a very good approximation. Then the ratio of
the CP conjugate amplitudes is
\beq\label{Dratio}
{A(D^0 \to \Ksm K^+) \over A(\D0bar \to \Ksm K^+)}
  = r_D\, e^{i \delta_D}\,.
\eeq

With these definitions, the four amplitudes we are interested in are given by
\beqa\label{4decays}
A[B^- \to K^- (\Ksp K^-)_D] &=& |A_B A_D|\, 
  \Big[1+r_B \, r_D\, e^{i(\delta_B+\delta_D-\gamma)} \Big], \nn\\
A[B^- \to K^- (\Ksm K^+)_D] &=& |A_B A_D|\, e^{i\delta_D} 
  \Big[ r_D + r_B e^{i(\delta_B-\delta_D-\gamma)} \Big], \nn\\
A[B^+ \to K^+ (\Ksm K^+)_D] &=& |A_B A_D|\, 
  \Big[ 1 + r_B\, r_D\, e^{i(\delta_B+\delta_D+\gamma)} \Big], \nn\\
A[B^+ \to K^+ (\Ksp K^-)_D] &=& |A_B A_D|\, e^{i\delta_D} 
  \Big[ r_D + r_B\, e^{i(\delta_B-\delta_D+\gamma)} \Big].
\eeqa
Of the unknowns in these equations, $|A_D|$ and $r_D$ have been measured in $D$
decays~\cite{PDG}, and $|A_B|$ was measured from the $B^- \to D^0 K^-$ rate
(and its conjugate) by reconstructing the $D^0$ in flavor-specific
decays~\cite{DK-meas,DKs-meas,DCP-K}.  (While in practice, measuring $|A_B|$
involves identifying the $D^0$ through its hadronic decay, which is not a  pure
flavor tag, this induces a negligible error.)  For any given integrated
luminosity in the future, the errors in the measurements of $|A_D|$, $r_D$, and
$|A_B|$ will induce a smaller error in the measurement of $\gamma$ than the
statistical error of measuring the decay rates corresponding to
Eqs.~(\ref{4decays}).

This brings us to the key point: by measuring the rates of the four decays in
Eqs.~(\ref{4decays}), one has four measurements for the remaining four
unknowns: $r_B$, $\delta_B$, $\delta_D$, and $\gamma$. 

A simple analytic solution for $\gamma$  can only be obtained neglecting the
terms proportional to $r_B^2$ in the four decay rates corresponding to the
amplitudes of Eqs.~(\ref{4decays}).  In this approximation, we obtain
\beq\label{solution}
\cos^2\gamma = {(R_1 + R_3 - 2)^2 - (R_2 + R_4 - 2r_D^2)^2 \over
  4\, [(R_1 - 1) (R_3 - 1) - (R_2 - r_D^2) (R_4 - r_D^2)] } \,,
\eeq
where 
\beq
R_1 = \left| {A[B^- \to K^- (\Ksp K^-)_D] \over A_B A_D} \right|^2 ,
\eeq
and similarly $R_{2-4}$ are the squares of the ``reduced amplitudes'' 
corresponding to lines $2-4$ in Eqs.~(\ref{4decays}).  Eq.~(\ref{solution})
illustrates that our method is sensitive to $\gamma$ even neglecting terms in
the branching ratios proportional to $r_B^2$.  Although $r_B$ is expected to
be small, $r_B^2$ should of course not be neglected when the experimental
analysis is carried out numerically.  On the other hand, $r_D$ is expected to
be of order unity, and our method works best if $r_D$ and $1-r_D$ are both of
order unity.  This expectation is supported by the data, $r_D = 0.73 \pm
0.21$~\cite{PDG} (the real uncertainty of $r_D$ may already be smaller; here we
assumed that the errors of the measured $D^0\to \Ksp K^-$ and $D^0\to \Ksm K^+$
rates are uncorrelated).

In principle, our method requires the analysis of only one type of final state
with different charge assignments (such as $K^{*\pm} K^{\mp}$).  In practice,
the sensitivity can be improved by considering several $B$ and $D$ decays of
the type considered so far. When a different $B$ decay mode is used, for
example, $B^- \to K^{*-} (\Ksp K^-)_D$, four more measurements can be done, but
only two new parameters are introduced, $r'_B$ and $\delta'_B$.  (Here we
assumed, as before, that $|A_B'|$ of this $B$ decay channel is measured.)  
When a different $D$ decay mode is used, for example, $B^- \to K^- (\rho^+
\pi^-)_D$, only one new parameter is introduced,  $\delta'_D$.  (Here we
assumed again that $|A_D'|$ and $r_D'$ of this $D$ decay channel are
measured.)  An especially interesting additional $D$ decay mode is to CP
eigenstates.  In this case, two extra measurements are possible, but no new
parameters are added, since $r'_D = 1$ and $\delta'_D = 0$ or $\pi$.

Next we discuss how the sensitivity to $\gamma$ changes in some limiting
cases.  If $r_D = 1$ then Eqs.~(\ref{4decays}) become degenerate if either
$\delta_D = 0$ or $\delta_B = 0$, and $\gamma$ can no longer be extracted from
this mode alone.  If $r_D \neq 1$, then our method is sensitive to $\gamma$
independent of $\delta_D$ and $\delta_B$.  However, if $\delta_D = 0$ or
$\delta_B = 0$ then the sensitivity to $\gamma$ comes only from terms in the
branching ratios proportional to $r_B^2$.  In this case Eqs.~(\ref{4decays})
are not degenerate, but both the numerator and the denominator of
Eq.~(\ref{solution}), obtained neglecting $r_B^2$ terms, vanish due to
$|\cos(\delta_D+\delta_B)| = |\cos(\delta_D-\delta_B)|$.  This indicates that
the error in the determination of $\gamma$ may become large if either of the
strong phases is small.  This potential difficulty may be eliminated using
several decay modes, as long as there are two sizable and different strong
phases.  For example, even if $\delta_B$ is small, the sensitivity of our
analysis to $\gamma$ can still be large and not rely on the $r_B^2$ terms in
the decay rates if we use two $D$ decay modes with sizable strong phases.

Throughout this analysis we assumed that $|A_D|$ and $r_D$ are known from $D$
decays, but $\delta_D$ is not. In the near future, it will be possible to
measure $\delta_D$ at a charm factory using CP-tagged $D$
decays~\cite{DDbar1,GGR}, simplifying our $\gamma$ measurement. With some model
dependence, $\delta_D$ may also be measured at the $B$ factories using a Dalitz
plot analysis of the $D$ decay~\cite{Williams}.  In this case, one typically
assumes that the variation of $\delta_D$ over the Dalitz plot can be accounted
for by using phases of Breit-Wigner resonances. With enough events, the
validity of this assumption can be checked in the analysis.  Both the charm and
the $B$ factory measurements of $r_D$ and $\delta_D$ can be carried out as a
function of the Dalitz plot variables.

Measurements of $\gamma$ that depend only on one strong phase, $\delta$, are in
general subject to an eight-fold discrete ambiguity, due to invariance of the
observables under the three symmetry operations~\cite{soffer}
\begin{equation}
\begin{array}{llll}
\sex &:~& \gamma \rightarrow \delta\,, & \delta \rightarrow \gamma\,, \\
\ssign &:~&\gamma \rightarrow -\gamma\,, & \delta \rightarrow -\delta\,, \\
\spi &:~& \gamma \rightarrow \gamma + \pi\,, \ \
  & \delta \rightarrow \delta + \pi\,.
\label{eq:ambig}
\end{array}
\end{equation}
In the modes we propose, the variation of $\delta_D$ across the $D$ decay
Dalitz plot will be largely determined by the Breit-Wigner shape of the
dominant resonance (such as the $K^*$). As a result, the theoretical 
expressions for the decay rates are no longer invariant under $\sex$ and
$\ssign$, and the only ambiguity that is relevant for our method is the
two-fold $\spi$ ambiguity~\cite{APS}.

To compare the sensitivity to $\gamma$ of the different methods, we assume that
the only small parameters are $\lambda$ and $r_B$.  The latter has been
estimated assuming factorization as
\beq
r_B \sim \left|{V_{ub} V_{cs}^*\over V_{cb} V_{us}^*}\right| {1\over N_C}
  \sim \lambda / 2\,,
\eeq
where $N_C = 3$ is the number of colors.  The accuracy of this estimate is
expected to depend on the specific hadronic mode.  Thus, for example, the
numerical values of $r_B$ in $B \to K (K^* K)_D$ and in $B \to K^* (K^* K)_D$
are expected to be different.  Since the uncertainty of this estimate of $r_B$
is large, it is not yet known whether $r_B$ is closer to $\lambda$ or to
$\lambda^2$.  The statistical significance of a CP asymmetry measurement scales
roughly as the smallest amplitude that is needed in order to generate the
asymmetry.  Thus, to compare the methods we need to identify the smallest such
amplitude in each of them.  Compared to $A_B$, the smallest amplitude in the
ADS method is of order $\min(\lambda^2,r_B)$, while in the GW and in our
methods it is of order $r_B\lambda$. This simple argument suggests that if $r_B
<\lambda$ then the ADS method has the largest sensitivity while if $r_B >
\lambda$ then it is one of the others.  There are many additional factors and
experimental differences that will influence this comparison when the
measurements are actually carried out.  For example, since in our case
measurements of doubly Cabibbo-suppressed $D$ decays are not needed, the
induced experimental error from $D$ decay rates is expected to be the
smallest.  We conclude that the sensitivity of these methods are comparable and
depend on yet unknown hadronic amplitudes and experimental details, and so all
should be pursued.

So far we have neglected $D-\Dbar$ mixing in our analysis. Its effects on the GW
and ADS methods were studied in Ref.~\cite{DDbar1}, and it is straightforward
to generalize it to our case.  We find that the leading effect on our
determination of $\gamma$ is generically of order $x_D/r_B$ and $y_D/r_B$,
where $x_D = \Delta m_D / \Gamma_D$ and $y_D = \Delta\Gamma_D / 2\Gamma_D$.
Since the present experimental bounds on $x_D$ and $y_D$ are at the few percent
level, and even the standard model values may be not much smaller~\cite{FGLP},
$D-\Dbar$ mixing gives rise to a theoretical error of order $10\%$.  This is
the largest theoretical uncertainty in our method at present, but it will be
reduced as experiments yield tighter  bounds on or measurements of $x_D$ and
$y_D$.  (The leading sensitivity to $D-\Dbar$ mixing in the GW method is
similar to our case, while in the ADS method there are potentially even larger
effects of order $x_D/\lambda^2$ and $y_D/\lambda^2$.)

We note that while there are experimental advantages to using the resonant $D$
decay final states $K^* K$ or $\rho\pi$, in general the full three-body Dalitz
plot may be used to perform this analysis. (Higher multiplicity final states
may also be used, although they will suffer from low reconstruction
efficiencies.) Our method should also work well in some regions of the Dalitz
plots of Cabibbo-allowed decay modes, such as $K_S \pi^+ \pi^-$ or $K_S K^+
K^-$. However, the regions where $r_D$ and $1-r_D$ are both of order unity are
expected to be relatively small, and so the advantage of the high branching
fractions of these dacays is not fully realized.

In conclusion, we proposed a variant of the GW method to measure $\gamma$. 
It requires measurement of only Cabibbo-allowed and singly Cabibbo-suppressed
$D$ decays in color allowed $B$ decays.  Because it involves only large decay
rates it may be carried out with current data sets.  The branching fraction and
reconstruction efficiency of the decay $B^\pm \to K^\pm (K K^*)_D$ is similar
to that of $B^\pm \to K^\pm (K^+ K^-)_D$, which has already been
observed~\cite{DCP-K}.  This might provide the first measurement of $\gamma$
that is free of hadronic uncertainties.

\begin{acknowledgments}

Y.G.\ was supported in part by the Israel Science Foundation
under Grant No.~237/01.
Z.L.\ was supported in part by the Director, Office of Science, Office of High
Energy and Nuclear Physics, Division of High Energy Physics, of the U.S.\
Department of Energy under Contract DE-AC03-76SF00098 and by a DOE Outstanding
Junior Investigator award.
The work of Y.G.\ and Z.L.\ was also supported in part by the United
States--Israel Binational Science Foundation through Grant No.~2000133.
The work of A.S.\ was supported by the
U.S. Department of Energy under contract DE-FG03-93ER40788.

\end{acknowledgments}

\end{document}